\documentclass[
a4paper,
prd,
twocolumn,
superscriptaddress,
preprintnumbers,
nofootinbib,
nobibnotes,
amsmath,amssymb,
aps,
floatfix
]{revtex4-2}
\usepackage{amsfonts,graphics,color}
\usepackage[english]{babel}
\usepackage{graphicx}
\usepackage{amsmath}
\usepackage{dcolumn}
\usepackage{bm}
\usepackage{multirow}
\usepackage{xcolor}
\usepackage{color}
\usepackage{comment}
\definecolor{xlinkcolor}{cmyk}{1,1,0,0}
\usepackage[bookmarks=true, pdfnewwindow=true, colorlinks=true, linkcolor=xlinkcolor, citecolor=xlinkcolor, filecolor=xlinkcolor, urlcolor=xlinkcolor, final=true]{hyperref}

\usepackage[normalem]{ulem}

\begin{document}

\preprint{INR-TH-2025-019}

\title{\texorpdfstring{Resonant production of millicharged scalars in $k^2 >0$ electromagnetic wave background}{Resonant production of millicharged scalars in k2>0 electromagnetic wave background}}

\author{Ekaterina Dmitrieva}
\email[\textbf{e-mail}: ]{edmitrieva@inr.ru}
\thanks{corresponding author}
\affiliation{Institute for Nuclear
Research of the Russian Academy of Sciences, 60th October Anniversary Prospect 7a, Moscow 117312, Russia}
\affiliation{Faculty of Physics, Moscow State University, Leninskiye Gory, 119991 Moscow, Russia}

\author{Petr Satunin}
\email[\textbf{e-mail}: ]{satunin@ms2.inr.ac.ru}
\affiliation{Institute for Nuclear
Research of the Russian Academy of Sciences, 60th October Anniversary Prospect 7a, Moscow 117312, Russia}
\affiliation{Faculty of Physics, Moscow State University, Leninskiye Gory, 119991 Moscow, Russia}
\affiliation{Branch of Lomonosov Moscow State University in Sarov, Parkovaya 2, 607328 Sarov, Russia}
\date{September 2026}

\begin{abstract}

    We investigate the solution of the Klein-Gordon equation for a charged scalar particle in an electromagnetic plane wave background with $k^2>0$, which can be realized in a medium with a refractive index $n<1$.
    We reduce the equation of motion to the Mathieu equation, which has resonant solutions that grow exponentially for certain parameter ranges.
    Physically, this effect leads to anomalous losses for certain beams of electromagnetic fields in a given medium. We set the constraints on the coupling and mass of millicharged particles from the absence of such anomalous losses for optical laser and radio beams.

\end{abstract}

\maketitle

\section{Introduction}

Particle production in an intense external field is still an intriguing phenomenon. The presence of the external field modifies the wave function of outgoing particles, allowing for some processes that are forbidden in standard perturbation theory based on the plane wave decomposition. A famous example is the Schwinger effect -- electron-positron pair production in an external electric field  \cite{PhysRev.82.664} and related processes (see books and reviews \cite{Ritus, Dittrich:2000zu, Kuznetsov:2013sea, Fedotov:2022ely}). Another nonperturbative effect related  only to the case of boson production  is the Bose enhancement: the production rate resonantly grows if final particles are produced in the background of  already occupied states of the same field.

This effect appears in cosmology as the so-called preheating after inflation, when a coherent oscillating heavy inflaton field resonantly decays to the matter scalars \cite{Kofman:1994rk,Kofman:1997yn,Khlebnikov:1996mc,Khlebnikov:1996wr} (see also the book \cite{lozanov2020reheating}). This decay process continues even when the produced scalars occupy all momentum states at that stage of the Early Universe; thus, the Bose enhancement effect becomes crucial. Mathematically, the equation of motion for produced scalar modes of fixed wavevector $\vec{k}$ reduces to the Mathieu equation, which parametric plane has bands of periodic and exponentially growing solutions; see the book \cite{mclachlan1947theory}. The resonance is of a classical nature; the quantum properties are not crucial.

The similar effect of resonance in particle production appears when we consider an initial state field configuration as an intense beam or a plane wave of fixed spatial momentum instead of the previously considered oscillating field \cite{Arza:2020zop, our}. If the plane wave is massive, the task reduces to the previous one by a relativistic boost; for an initial massless plane wave, the resonant effect still occurs in a toy model with two scalar fields, described in this case by tachyonic instability \cite{Arza:2020zop, our}. In both cases, the resonant exponential growth stops, and the energy density of the produced field becomes of the same order as that of the initial field.  Unlike the preheating that has occurred everywhere in the Early Universe, a physical beam  of any field must be of finite width; if a final particle produced in the central part of the beam reaches its border, the resonance stops as well.       


In this article, we explore the generalization of the aforementioned mechanism to particle production in an electromagnetic plane wave background. It is well-known that a plane wave of the electromagnetic field in vacuum (the squared 4-momentum vanishes, $k^2=0$) is stable: there is no tachyonic instability or Schwinger effect in such a field configuration \cite{PhysRev.82.664}.  Besides the plane wave in a vacuum, one can consider plane waves in different types of media, obtaining both $k^2<0$ in standard crystals and $k^2 > 0$ in thermal plasma or in some types of metamaterials  \cite{ Wood_2007, PhysRevLett.102.093903, navau2009magnetic, magnus2008dc, PhysRevB.77.092401, PhysRevB.87.024408}.  
In $k^2>0$ background, the photon becomes effectively massive; in the presence of charged particles with small masses in the theory, one expects a similar type of particle production as for scalars \cite{our}. 

Such particles with tiny electric charges are actively discussed as  one of the ``new physics'' extensions of the Standard model and are called millicharged particles (mCP) 
\cite{IGNATIEV1979315, Holdom:1985ag,Holdom:1985ag, Batell:2005wa, Abel:2003ue, Abel:2008ai}. 
Besides, mCP emerges in cosmology as a candidate for the dark matter particle, either alone or as  part of wider dark sector models \cite{Brahm:1989jh, Bogorad:2021uew, Jaeckel:2021xyo, Shiu:2013wxa, Feng:2009mn, Cline:2012is, Okun:1982xi, Holdom:1985ag}.
Searches for these particles are conducted in various experiments \cite{Berlin:2024dwg, Arza:2025cou, Gninenko:2006fi, SENSEI:2023gie, deMontigny:2023qft, Berlin:2025btf, Berlin:2025hjs, Gninenko:2025aek} and astrophysical observations \cite{Fiorillo:2024upk, Davidson:2000hf, Berlin:2021kcm, Chang:2018rso,Lepidi:2007vnd, Cruz:2022otv}.

The wave function for relativistic scalar electrons in a medium with nontrivial $n$ has been studied in \cite{CRONSTROM1977137,BECKER1977601,Heinzl:2016kzb, King:2016oei}. It was shown that the equation for the wave function reduces to the Mathieu equation, whose solutions are finite oscillating  or exponentially growing in different bands of the parametric plane. 
In this article, we interpret this growing solution as the solution to the Klein-Gordon equation for the classical mCP field in the electromagnetic wave background, describing the resonant exponential growth of mCPs.

\begin{figure}
    \centering
    \includegraphics[width=1.0\linewidth]{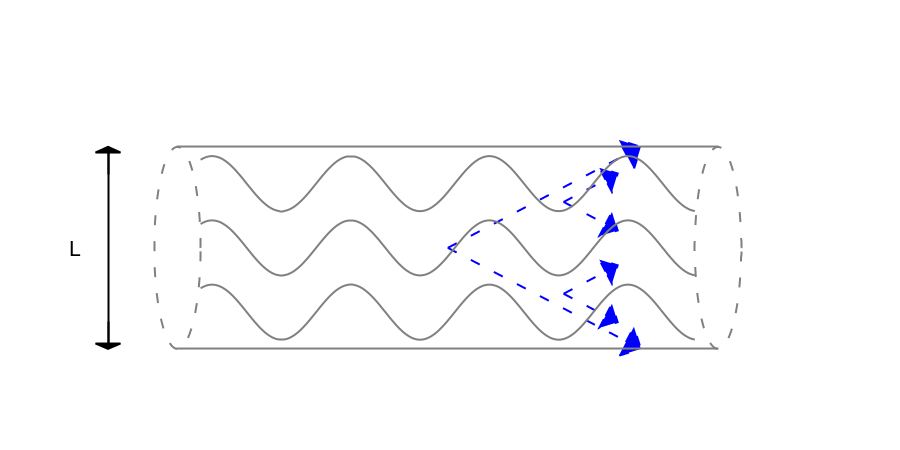}
    \caption{An idea of resonance growth of millicharged scalars in an electromagnetic coherent beam. A pair of mCP modes of opposite transverse momenta produces inside the beam; in the process of this propagation to the border of the beam, the amplitude of the mode coherently superimposed to the same mode produced at the trajectory of the mode. The Bose enhancement effect stimulate the resonance growth.
    }
    \label{fig:picture_coh}
\end{figure}

 In our setup, the external field is modeled as a coherent plane wave of finite transverse size. 
 As the mCPs created in the central part of the beam propagate towards the beam periphery, Bose enhancement amplifies the production rate, leading to a significant increase in their number. This process continues until the mCPs exit the beam. This process is illustrated in figure \ref{fig:picture_coh}, where $L$ is the beam width. At the same time, we restrict our analysis to the initial stage of the interaction, during which the total number of produced particles remains well below the photon population in the beam. This assumption justifies neglecting back reaction effects. Subsequently, we apply our formalism to experimentally relevant sources — a laser and a radio wave — explicitly accounting for the finite transverse dimensions of the beams.

The paper is organized as follows. In Sec. \ref{sec_Mathieu} we consider the Klein-Gordon equation for a running wave and reduce it to the Mathieu equation. In sections \ref{sec_narrow} and \ref{sec_broad}, we investigate instability bounds from narrow and broad resonance, respectively. In Sec. \ref{sec_standing_wave} we investigate the standing wave and compare the results of this section with experimental constraints. In Sec. \ref{sec_conclusion} we make a conclusion.

\section{\texorpdfstring{Particle production in running wave with $k^2>0$ background}{Particle production in running wave with k2>0 background}}
\label{sec_Mathieu}

We consider the production of a charged scalar field in an external field of a running electromagnetic wave in a medium, described by the dispersion relation $\omega = n^{-1} \cdot |\vec{k}|$, where $(\omega, \vec{k})$ is the momentum four-vector, and $n$ is the medium refractive index. The Lagrangian for the scalar field $\psi$ reads,
 \begin{equation}
    \mathcal{L} = D_\mu\psi^* D_\mu \psi - m^2\psi^*\psi, 
\end{equation} 
where $D_\mu = \partial_\mu + ie\mathcal{A}_\mu $, where $\mathcal{A}_\mu$ is electromagnetic potential. The classical dynamics of the scalar field $\psi$ are governed by the equation of motion, which is the Klein-Gordon equation
\begin{equation}\label{K-G}
    (\Box+2ie\mathcal{A}^\mu\partial_\mu-e^2\mathcal{A}^2+m^2)\psi=0,
\end{equation}
where we take into account the gauge $\partial_{\mu}\mathcal{A}^{\mu}=0.$
Notably, the same equation is obtained in quantum mechanics (QM) as the relativistic extension of the Schrödinger equation for the wave function of a charged scalar particle \cite{CRONSTROM1977137,BECKER1977601}; the difference is that the QM wave function should be normalized to unity;  there is no such condition for the scalar field.

We are looking for the solution of eq.~(\ref{K-G}) using the following ansatz,
\begin{equation}
\label{a1}
\psi=\psi_p=e^{-ip\cdot x}F_p(\phi),
\end{equation}
where $\phi\equiv k\cdot x$ (here and further, the notation $'\cdot \,'$ means the scalar product of four vectors), and $p_\mu$ is an arbitrary four vector that satisfies the condition $p^2=m^2$. The index $p$ represents the momentum of the resulting particles. Note, that there is a symmetry regarding the replacement of the pulse $p$ by $k-p$, see the Appendix \ref{app_k-p} for details. 
Inserting the ansatz \eqref{a1} into the equation \eqref{K-G}, we obtain the expression
\begin{equation}
\label{runningwave}
    k^2F_{p}''(\phi)-2i\,p\cdot kF_{p}'(\phi)+(2e\,p\cdot \mathcal{A}-e^2\mathcal{A}^2)F_{p}(\phi)=0.
\end{equation}
We consider a circularly polarized monochromatic wave,
\begin{equation}\label{eq_polariz}
\mathcal{A}^{\mu}=a^{\mu}\cos\phi+b^{\mu}\sin\phi,~~~a\cdot a=b\cdot b<0,
\end{equation}
where the momentum $\vec{k}$ is directed along the $z$-axis; 
\begin{equation}
\label{eq:ab}
a_\mu=(0,E_0/\omega,0,0)\qquad \text{and}\qquad b_\mu=(0,0,E_0/\omega,0) 
\end{equation}
there are two constant space-like orthogonal four-vectors determining the polarization, $E_0$ is the amplitude of the electromagnetic wave. We introduce the abbreviations:
\begin{equation}
 \phi_0 = \arctan (p \cdot b/p \cdot a), \quad \zeta = \frac{1}{2}\left( \phi - \phi_0\right).
\end{equation}
Using the subsequent ansatz 
\begin{equation}
F_{p}(\phi)=w_{p}(\zeta)\mbox{exp}\left[i\left(\frac{(p\cdot k)}{k^2}\right)\phi\right],
\end{equation}
we finally reduce the equation of motion (\ref{K-G}) to the form of the Mathieu equation, cl. \cite{CRONSTROM1977137, BECKER1977601}
\begin{equation}\label{Mathieu}
    \frac{d^2w_{p}}{d\zeta^2}+(A_p+2q\cos(2\zeta))w_{p}=0,
\end{equation}
where 
\begin{equation}
\label{eq:Ap}
   A_p=\frac{4}{k^2}\left(\frac{(p\cdot k)^2}{k^2}-e^2a^2  \right),
\end{equation}
and
\begin{equation}
\label{eq:q}
    q=\frac{4e}{k^2}((p\cdot a)^2+(p\cdot b)^2)^{1/2}.
\end{equation}



\begin{figure}
    \centering
    \includegraphics[width=0.99\linewidth]{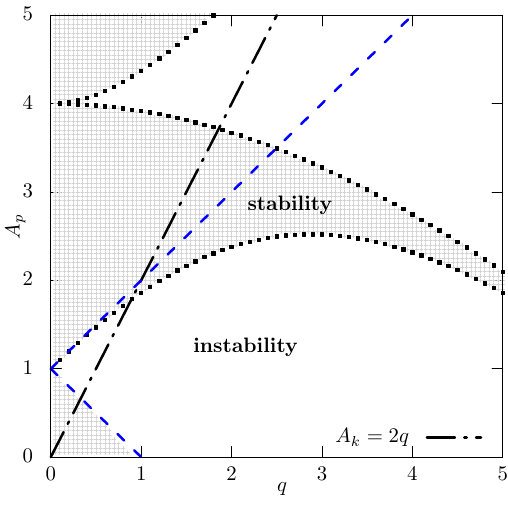}
    \caption{Mathieu equation stability chart. Vertical axis: $A_p$, Horizontal: $q$. Grey(white) area: stability(instability) regions of Mathieu equation. Blue dashed line: Instability bound for narrow resonance regime.  
    Black line of dots and dashes - line $A_p=2q$.} 
    \label{fig_Mathieu_stab/inst_Arza}
\end{figure}

The properties and solutions of the Mathieu equation (\ref{Mathieu}) are well-studied; see, e.g., the book \cite{mclachlan1947theory}.
The main property of the Mathieu equation solution is that, depending on the values of the parameters $(A_p,q)$, the solutions are periodic (stable) or exponentially growing (unstable). The corresponding areas of stability and instability are presented in the Mathieu equation chart; see Fig.~\ref{fig_Mathieu_stab/inst_Arza}. The instability areas, or the resonant bands, are the stripes that start for $q=0$ at the points $A_p = N^2$, where $N$ is a natural number, and grow wider with $q$.

Depending on the values of $A_p$ and $q$, one distinguishes two regimes  of analytical solutions \cite{Kofman:1997yn}: narrow resonance, related to the case $q\ll 1$, $A_p \sim 1$ (the beginning of the first resonant band at Fig.~\ref{fig_Mathieu_stab/inst_Arza}), and broad resonance ($A_p \sim 2q \gg 1$); both of which are studied below in detail. 

Physically, the process can be interpreted as the production of pairs of charged scalar particles enhanced by parametric resonance. The creation of a particle with momentum $p$ is described by the exponentially growing solution of the Mathieu equation \eqref{Mathieu}. The second particle from the pair corresponds to the solution for the conjugate mode with momentum $k-p$, which is analyzed in detail in Appendix \ref{app_k-p}. As shown there, the Mathieu equation for that mode takes the same form as \eqref{Mathieu} with the substitution $p\to-p$. Furthermore, in the narrow‑resonance regime, the corresponding Floquet exponent $\mu_{Fl}$ (see details below) exhibits identical growth. This confirms the interpretation of the process as pair production of two scalar particles. Note that during this process, many pairs of particle-antiparticle with momenta $p$ and $k-p$ are generated.

The plane wave is not a solution of EOM \eqref{K-G}. However, the plane wave case is the limiting case; see perturbative regime.

For simplicity, we begin with the perturbative regime, which corresponds to the narrow-resonance limit $q\to 0$.

\subsection{Perturbative regime. }

Perturbatively, a photon of 4-momentum $k_\mu=(\omega,0,0,n\omega)$ decays to a pair of charged massive scalars of, for simplicity, the same energy and the same $z-$projection of spatial momentum,
\begin{equation}
\label{pmu}
  p_\mu =  \left(\sqrt{\left(\frac{n\omega}{2}\right)^2+m^2+p^2_1+p^2_2},\pm p_1,\pm p_2,\frac{n\omega}{2}\right).
\end{equation}

Here, we consider $p_z=n\omega/2$ because, in this case, we obtain the maximal instability area and also, for illustration, and simplify the calculation.
The form (\ref{pmu}) satisfies $p^2=m^2$. The energy-momentum conservation implies the value for the perpendicular momentum for finite particles, 
\begin{equation}
\label{pperp}
     p_\perp^2  \equiv p_1^2 + p_2^2  = \frac{\omega^2(1-n^2)}{4} - m^2.
\end{equation}
$p^2_\perp$ is positive for a medium with $n < 1$ and for a quite small mass $m$ of a millicharged scalar. Substituting \eqref{pperp} to \eqref{eq:Ap} we obtain $A_p = 1$. A single initial photon instead of a classical wave represents the limiting case $q \to 0$.  We expect that the narrow resonance is an extension of this process to larger amplitude $E_0$ and is more efficient due to Bose-enhancement.

The width of the perturbative process can be reduced to the photon decay calculation with Lorentz invariance violation at the fundamental level \cite{Rubtsov:2012kb}, which gives, in our terms,
\begin{equation}
    \Gamma_{pert} = \frac{e^2\omega}{24\pi}\left( \frac{1}{n^2} - 1\right).
\end{equation}

\subsection{Narrow resonance. First resonance band.}\label{sec_narrow}

The parametric resonance occurs when the solution increases exponentially over time. For the Mathieu equation, this case corresponds to unstable solutions.

Now we consider the nonzero classical field amplitude $E_0$. We keep the condition $q \ll 1$. This condition corresponds to the narrow resonance regime, which has the following physical explanation: the perturbative process of producing particles and narrow resonance enhances the effectiveness of this process. We make the following dimensionless notations: 
\begin{equation}
\label{eq:Emupi}
    \mathcal{E} \equiv \frac{eE_0}{\omega^2}, \qquad \mu \equiv \frac{m}{\omega}, \qquad \pi_\perp \equiv \frac{p_\perp}{\omega}.
\end{equation}
The expressions \eqref{eq:Ap} and \eqref{eq:q} on the momentum configuration \eqref{pmu} read,
\begin{equation}
\label{eq:Ap1}
   A_p=\frac{4\left(\sqrt{\frac{n^2}{4}+\mu^2+\pi^2_\perp}-\frac{n^2}{2}\right)^2}{(1-n^2)^2}\!+\!\frac{4\mathcal{E}^2}{1-n^2} ,
\end{equation}
and
\begin{equation}
\label{eq:q1}
    q=\frac{4\mathcal{E}}{1-n^2}\pi_\perp.
\end{equation}

The solutions in the unstable region grow exponentially $w \sim e^{\mu_{Fl}\zeta}$ \cite{mclachlan1947theory}, see Appendix \ref{app_Fl} 
for the exact form of the solutions, with the Floquet exponent  $\mu_{Fl}$ determined in the region $q \ll 1$ as \cite{mclachlan1947theory}
\begin{equation}
\label{eq:MuFl}
    \mu_{Fl} = \frac{1}{2}\sqrt{q^2 - (A_p-1)^2}.
\end{equation}
The boundary between the stable and unstable regions of the Mathieu chart (Fig.\ref{fig_Mathieu_stab/inst_Arza}) is $\mu_{Fl}=0$, or
\begin{equation}
\label{eq:NarrowBound}
    q = \left| A_p - 1\right|.
\end{equation}
This dependence is shown in Figure \ref{fig_Mathieu_stab/inst_Arza} by the dashed blue line.

\begin{figure}
    \centering
    \includegraphics[width=0.995\linewidth]{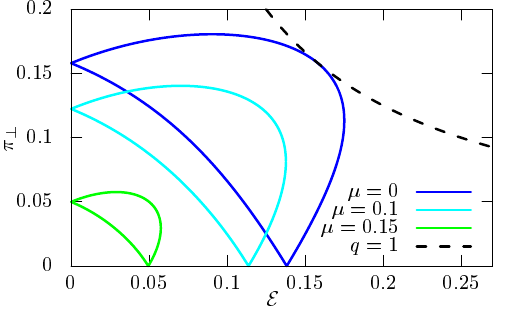}
    \caption{The instability bounds of scaled perpendicular momentum $\pi_\perp$ as a function of scaled electric field $\mathcal{E}$, $n=\sqrt{0.9}$. 3 curves: $\mu = 0$, $\mu=0.1$, $\mu=0.15$. The boundaries of instability region determined by eq.~\eqref{eq:MuFl} for $\mu_{Fl}=0$. The dashed curve represents $q=1$, the area of the narrow resonance regime is below the line and inside the areas bounded by solid lines.}
    \label{fig:3mu}
\end{figure}

The stability boundary \eqref{eq:NarrowBound} expressed through the model parameters, see \eqref{eq:Ap1} and \eqref{eq:q1}, can be solved with respect to the perpendicular momentum $\pi_\perp$. The numerical solution for the boundary $\pi_\perp$ as a function of the scaled amplitude $\mathcal{E}$ is shown in Fig.~\ref{fig:3mu} (numerical values $1-n^2=0.1$ and $\mu=(0,0.1,0.15)$ are taken). The instability area in the parametric plane (related to the narrow resonance) lies inside the figure formed by the branches of the solution. 

\begin{figure}[h]
    \centering
    \includegraphics[width=0.48\textwidth]{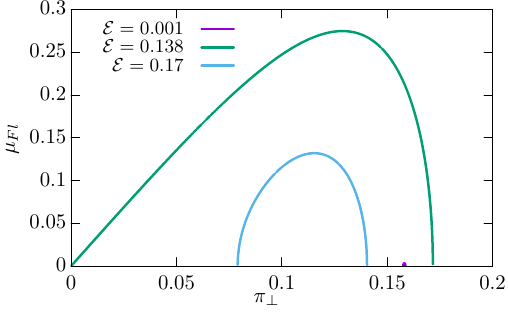}
    \caption{Floquet exponent $\mu_{Fl}$ dependence on the scaled perpendicular momentum $\pi_\perp$ for first narrow resonance band and fixed $n=\sqrt{0.9}$, $\mu=0$, for several values of scaled electric field $\mathcal{E}$.}
    \label{fig_mu_fl(p)_dif_eE}
\end{figure}

\begin{figure}[h]
     \includegraphics[width=0.48\textwidth]{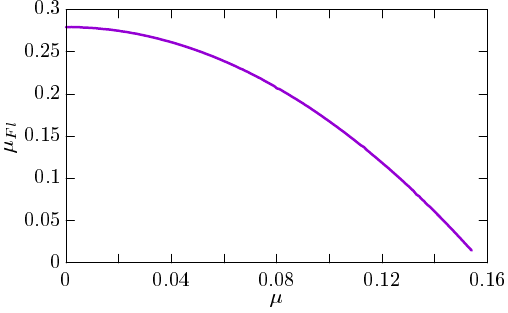}
    \caption{Dependence of $\mu_{Fl}$ on $\mathcal{E}$ for a fixed value of $\pi_\perp$ corresponding to the maximum of $\mu_Fl$.
    }\label{fig_mu_fl(mu)}   
\end{figure}

\begin{figure}[h]
     \includegraphics[width=0.48\textwidth]{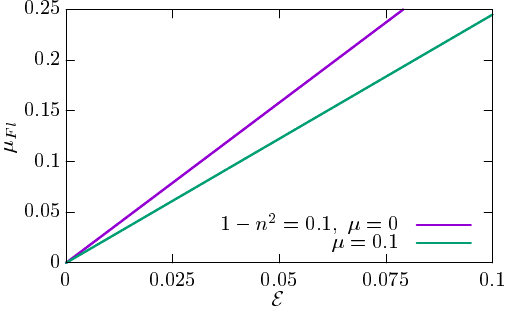}
    \caption{Floquet exponent $\mu_{Fl}$ dependence on scaled electric field $\mathcal{E}$ for first narrow resonance band and fixed $n=\sqrt{0.9}$ and fixed scaled mass $\mu=0, 0.1$ for approximate case $\mathcal{E}\ll1$ (see eq. \eqref{eq:MuFl_apr}).
    }\label{fig:muFl(E)_apr}   
\end{figure}


The dependence of the Floquet exponent $\mu_{Fl}$ (see eq.~\eqref{eq:MuFl}) on $\pi_\perp$ at fixed $\mathcal{E}$ is shown in Fig.~\ref{fig_mu_fl(p)_dif_eE}. $\mu_{Fl}$ almost disappears at $\mathcal{E} = 0.001$ and $\mathcal{E} = 0.175$. Analytically, it can be obtained that for $\pi_\perp = 0$ or for $\mathcal{E} = 0$, $\mu_{Fl}$ does not exist. From figure \ref{fig:3mu} we can see that for $\mu=0$ $\pi_\perp\approx0$ for $\mathcal{E}\approx0.138$. This correspond green line on figure \ref{fig_mu_fl(p)_dif_eE}. It can be concluded that the solution exists at $0.001\leq \mathcal{E}\leq 0.175$. These boundaries are approximate values for which $\mu>0$ and $\mu\to0$ are always positive, approaching zero $\mu$ within these values. These values are approximate and are obtained from graphs.


Moreover, in order to draw a graph of the dependence of $\mu_{Fl}$ on $\mu$, for each fixed $\mu$ we will find the dependence of $\mu_{Fl}$ on $\pi_\perp$ for various $\mathcal{E}$ and choose the one for which the area under the graph is maximal. For this $\mathcal{E}$, we will find the maximum point of $\pi_\perp$. Using this data for each $\mu$, we will plot the dependence of $\mu_{Fl}$ on $\mu$; it is shown in the figure \ref{fig_mu_fl(mu)}.

In the limit $\mathcal{E} \ll \sqrt{1-n^2-4\mu^2}$, eq.~\eqref{eq:NarrowBound} has an analytical solution with respect to $\pi_\perp$,
\begin{equation}
\label{pi_perp_small}
    \pi_\perp^\pm = \frac{1}{2}\sqrt{1-n^2-4\mu^2} \pm \mathcal{E}.
\end{equation}
The mCPs are produced resonantly for perpendicular momenta $ \pi_\perp^-< \pi_\perp< \pi_\perp^+$.

For small $\mathcal{E}$, we can use the central value of eq.~\eqref{pi_perp_small}, obtaining
\begin{equation}
\label{eq:MuFl_apr}
    \mu_{Fl} = \frac{\mathcal{E}\sqrt{1-n^2-4\mu^2}}{1-n^2}, \qquad \mathcal{E}\ll 1.
\end{equation}
The dependence of $\mu_{Fl}$ on the $\mathcal{E}$ for an average of the instable area value of $\pi_\perp$ (see eq. \eqref{pi_perp_small}) is shown in Fig.~\ref{fig:muFl(E)_apr}.

\subsection{Decay rate and mCP density} 
Now we consider the decay rate for the perturbative and resonance cases. 
The full photon decay rate to mCPs in a medium with $n > 1$, related to the perturbative one, is given by \cite{Arza:2020zop, Alonso-Alvarez:2019ssa, baumann2011physics},
\begin{equation*}
    \Gamma=\Gamma_{pert}(1+2N),
\end{equation*}
where $N$ is the occupation number for an mCP state. The decay rate is enhanced by a large occupation number $N$ for $N\gg1$, we reach the Bose enhancement regime: the second term yields exponential growth of the particle number. This is the narrow resonance. 

The Boltzmann equation for the mCP density reads \cite{Alonso-Alvarez:2019ssa},
\begin{equation}\label{eq_Boltzmann}
    \dot n_\psi=2\Gamma_{pert}(1+2n_\psi/V_{ph})n_\gamma;
\end{equation} 
here we take $N \equiv n_\psi / V_{ph}$, where $V_{ph}$ is the corresponding phase-space volume \cite{Alonso-Alvarez:2019ssa}. Here we will not calculate the phase volume, since it is not used in the final expressions.

The solution of eq.~\eqref{eq_Boltzmann} reads,
\begin{equation}
    n_\psi = \frac{V_{ph}}{2}\left(  e^{\Gamma_{pert}\cdot n_\gamma/V_{ph}\times (t-t_0)} - 1\right).
\end{equation}
there will be a perturbative limit for $N\ll1$. It corresponds to a small time of coupling, while the number of particles is small. Therefore, for small $(t-t_0)$, one obtains the perturbative result $\left.n_\psi\right|_{pert} = n_\gamma \times \Gamma_{pert}(t-t_0)/2$;  in the resonance regime, the exponent dominates, so we fit
\begin{equation}
    \Gamma_{pert}\cdot n_\gamma/V_{ph} =  \mu_{Fl} \omega.
\end{equation}
We obtain this expression using the following statements: unstable modes for the narrow resonance regime of the Mathieu equation are $\psi_k\sim e^{\mu_k z}$; therefore, occupation numbers correspond to $n_\psi\sim e^{2\mu_k z}=e^{2\mu_k\omega t}$ \cite{schmitz2010reheating, Kofman:1997yn}.
The narrow resonance regime is achieved only for large $N$, when particles are produced in the perturbative regime. The narrow resonance enhances the perturbative process.
Thus
\begin{equation}\label{eq_gamma_narrow}
   \left. n_\psi \right|_{res} = n_\gamma \frac{\Gamma_{pert}}{\mu_{Fl}\omega}  e^{\mu_{Fl}\omega\times (t-t_0)}.
\end{equation}
 and for $m=0$:

\begin{equation}
   \left. n_\psi \right|_{res} = n_\gamma \frac{e^2}{4\pi}\frac{(1-n^2)^{3/2}}{12 n^2\,\mathcal{E}}  e^{\frac{2\mathcal{E}\omega}{\sqrt{1-n^2}}\times (t-t_0)}.
\end{equation}



\subsection{Broad resonance.}\label{sec_broad}

For a narrow resonance, the condition $q \gtrsim 1$ is not applicable. Nevertheless, there exists another analytic solution, known as broad resonance, which operates in the regime $q \gg 1$ and is even more efficient. In the trilinear scalar models, this corresponds to non-perturbative limit.

A broad resonance occurs in bursts, each time the adiabaticity condition is violated, and an exponential increase in the number of particles occurs \cite{Kofman:1997yn, Kofman:1994rk,lozanov2020reheating}.

The broad resonance regime is characterized by the mCP undergoing many oscillations within a single oscillation period of the background field \cite{Kofman:1997yn, lozanov2020reheating}. In this case, the frequency of the background field is given by
\begin{equation} \label{omega_chi}
\Omega^2 = A_p + 2q\cos(2\zeta).
\end{equation}

In the broad resonance regime, particle production takes place in bursts. These bursts occur whenever the adiabatic condition is violated — that is, when the system departs from an adiabatic regime and the WKB approximation (see details in Appendix \ref{app_WKB}) ceases to be valid
\begin{equation}\label{adiab_cond}
    \frac{d\Omega}{d\zeta}\gtrsim\Omega^2.
\end{equation}
 For narrow resonance, this condition is not necessary. As a result, we obtain the criterion 
\begin{equation}\label{cond_broad}
(-2q\sin(2\zeta))^2\gtrsim\left(A_p+2q\cos(2\zeta)\right)^{3}.
\end{equation}

The parameter $\zeta$ corresponds to the time when the adiabatic condition is violated. At these moments in time, when the conditions of adiabaticity are violated, an exponential increase in the number of particles begins.

The main condition for broad resonance is $q\gg1$.  

From condition \eqref{cond_broad} for different values of $2\zeta$, we can obtain different conditions.
For $\zeta\to\pi+2\pi N, N\in\mathbb{Z}$, and near these values, we obtain
\begin{equation}\label{cond_broad_pi}
    A_p\lesssim2q.
\end{equation}

Here we are using the limits $q\gg1$ and $\sin(2\zeta)\approx0$ and $\cos(2\zeta)\approx1$.

The line $A_p=2q$ is shown in Figure \ref{fig_Mathieu_stab/inst_Arza} as a black dashed line. From this figure, we can see that this line is a rough bound for broad resonance. 
This boundary condition for dimensionless physical parameters acquires the following form:
\begin{equation}\label{cond_broad_dimensionless_rough}
\frac{\left(\sqrt{n^2/4+\mu^2+\pi^2_\perp}-n^2/2\right)^2}{(1-n^2)^2}+\frac{\mathcal{E}^2}{1-n^2}=\frac{2\mathcal{E}\pi_\perp}{1-n^2}.
\end{equation}  

For the refractive index $n<1$ the solution of \eqref{cond_broad} exists only for very small range of parameter $\zeta$. Moreover, solution in this range not satisfying condition $q\gg1$. Therefore the broad resonance for this refractive index does not exist.

\subsection{mCP propagation in an electromagnetic field}

The theory presented above can be extended to millicharged particles (mCPs). 

Once produced in a perturbative or resonant process, mCP evolves in the electromagnetic field. The important thing is that a mCP can quickly escape the region of the intensive field and, by that, stop the resonance.

The EOM for a  single relativistic millicharged particle reads 
\cite{Heinzl:2008rh},
\begin{equation}\label{traj_eom}
   \frac{dP^\mu}{d t}=\frac{e}{m}F^{\mu\nu}P_\nu,  
 \end{equation}
where $P_\mu$ is the particle 4-momentum, and $F_{\mu\nu}\equiv\partial_\mu A_\nu-\partial_\nu A_\mu$ is the electromagnetic field tensor.
 
 The particle trajectory in the plane perpendicular to the electromagnetic wave momentum can be approximated as  \cite{Raicher:2025kpl, Lv_2021}
\begin{equation}\label{trajectory_r}
x_\perp(t) = \frac{p_\perp}{\omega} t+\frac{eE_0}{\omega (k\cdot p)}\sin(\phi),
 \end{equation}
where $\phi\equiv k\cdot x$. 
 Let us call  $d$ the transverse dimension of the electromagnetic beam. The resonance stops if  $x_\perp(t) > d$ occurs, because particles escape the beam. Eq.~\eqref{trajectory_r} shows two conditions: first, the amplitude of the oscillating term in \eqref{trajectory_r} should be less than $d$;
 \begin{equation}
     \frac{eE_0}{\omega (k\cdot p)} < d.
 \end{equation}
 Second, the first term in Eq.~\eqref{trajectory_r} grows linearly with time. To achieve effective resonance,  the time for mCP to leave the beam should be much longer than the oscillation period;
\begin{equation}\label{traj_t}
    t \equiv \frac{d}{\pi_\perp} \gg \omega^{-1}.
\end{equation}

Furthermore, we check these conditions for concrete physical setups.

\subsection{Towards experimental challenges}\label{sec_exp}

\begin{figure}
    \centering
    \includegraphics[width=0.5\textwidth]{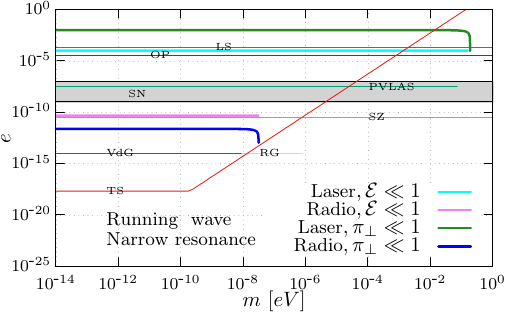}
    \caption{The projected bounds on charge $e$ and mass $m$ of mCP from the absence of anomalous losses of running electromagnetic wave (IR laser/Radio) in a metamaterial with refractive index $n=\sqrt{0.9}$, in comparison with existing bounds for mCP:  LS from Lamb shift \cite{Gluck:2007ia}, OP from Ortho-Positronium decay \cite{Badertscher:2006fm}, PVLAS from the PVLAS experiment \cite{DellaValle:2015xxa}, SN from SN1987A \cite{Chang:2018rso, Davidson:2000hf, Davidson:1993sj}, SZ from CMB observations of the Sunyaev-Zel'dovich effect \cite{Burrage:2009yz}, RG from red giants \cite{Davidson:1993sj}, VdG and TS from Van de Graaff and thunderstorms, respectively, for the mCP condensate \cite{Berlin:2024dwg, Dmitrieva:2026ges}.
    The green/blue curves show the narrow-resonance regime for radio and infrared (Nd:YAG) background waves, respectively, at fixed $\pi_\perp=0.01$. The cyan/violet curves show the same regimes for radio and infrared, respectively, with fixed $\mathcal{E}\ll1$.
    }
    \label{fig_e(m)_exp}
\end{figure}

In this subsection, we discuss the possibility of experimental tests of mCP parameters $(e,m)$ with the aforementioned resonant process of mCP production in an intense running wave of the electromagnetic field with dispersion $k^2>0$.  We concentrate on possible laboratory tests for which a metamaterial medium with $n<1$ is necessary.
The experimental signature of resonant mCP production can be provided in two ways: anomalous energy losses in the present electromagnetic wave or detection of produced mCPs by another process. Here, we adopt the first approach and neglect, for simplicity, the back reaction process, which becomes relevant after sufficient losses; see \cite{Arza:2020zop} for a relevant discussion. We consider only a short period of time, when the number of resulting particles is small, but fast increase and back reaction are irrelevant. 

A charge $e$ contributes only to parameter $\mathcal{E}$: $e=\frac{\omega^2}{E_0}\mathcal{E}$, while $m=\mu\cdot \omega$,  cl.~\eqref{eq:Emupi}.
We can obtain the dependence $e$ on $\mu$ from equation \eqref{eq_gamma_narrow}. We use this equation because we need to take into account that the narrow resonance is a continuation of the perturbative process and its amplification due to Bose-enhancement. We are using the condition that $n_{\psi}|_{res}=1/2\,n_{\gamma}$, because after that, the effects of back reaction became significant. Moreover, we must estimate $t$. We can do it in different ways. In Fig.\ref{fig:3mu} there are two different cases of $\mathcal{E}$ and $\pi_\perp$. The first case is the $\mathcal{E}\ll1$, which corresponds to equations \eqref{pi_perp_small} and \eqref{eq:MuFl_apr}. Also, we can estimate $t$ as eq.\eqref{traj_t}, where $\pi_\perp$ comes from eq.\eqref{pi_perp_small}. These expressions correspond to sufficiently large values of $\pi_\perp$. The second case is the $\pi_\perp\ll1$; we take a fixed value $\pi_\perp=0.01$. However, we cannot use the approximate equation \eqref{eq:MuFl_apr}; therefore, we will use eq. \eqref{eq:MuFl}.

Intuitively, the best sensitivity to $e$ is for a radio frequency wave. For this reason, we consider a benchmark wavelength $\lambda \sim 1$ m, which corresponds to a frequency $\omega \sim 3\times 10^5$ kHz and $\sim 2 \times 10^{-7}$ eV. We take the benchmark value for the amplitude of the electric field $E_0\sim 10\,\mbox{kV/m} \sim 2.3\times10^{-3}\, \mbox{eV}^2$ and $d\sim10\lambda$. The corresponding sensitivity for $e$ in the regime $\mathcal{E}\ll1$, $m \lesssim 3\times10^{-8}$ eV, can be $e \sim 4.5\times 10^{-11}$; see the solid violet line in Fig.~\ref{fig_e(m)_exp} (Radio). In the regime for $\pi_\perp \ll 1$ we obtain $m \lesssim 10^{-8}$ eV, can be $e \sim 2.2\times 10^{-12}$; see the solid blue line in Fig.~\ref{fig_e(m)_exp} (Radio).

Another experimental configuration that can potentially be effective is an optical or near-infrared laser. Instead of the case of radio waves, the width of the laser beam can be significantly larger than the wavelength; thus, the produced mCPs do not immediately escape the strong field region, avoiding resonance.  
We estimate the sensitivity for $e$ by taking the Nd-YAG laser ($\lambda \simeq 1 \;\mu$m), which is related to frequency $\omega\sim 1.2\,\mbox{eV}$; the field amplitude reads $E\sim10^8$ eV/m;  the beam width is $d=1\,\mbox{mm}$, which is $\sim 10^3$ wavelengths. 

The sensitivity in the regime $\mathcal{E}\ll1$ for $e$ reads $e \sim 9\times 10^{-5}$ and mass $m\lesssim0.2~\mbox{eV}$.
This sensitivity line is presented in Fig.~\ref{fig_e(m)_exp} (Laser) by a cyan line alongside a comparison with existing limits.
In the regime $\pi_\perp\ll1$ the charge estimates as $e \sim 2\times10^{-3}$ and mass $m\lesssim 0.2 ~\mbox{eV}$.

The next step is to consider a wave with an amplitude larger than previously. Therefore, we now proceed to analyze a standing wave in a cavity.

\section{Standing wave}\label{sec_standing_wave}

In addition to the running wave discussed in the previous section, we study resonant mCP production in a standing wave background. A standing wave in the radio frequency range can have a significantly larger amplitude than a running wave, as the electromagnetic energy is coherently accumulated within a closed volume. 

We model the standing wave as a superposition of two counter-propagating running waves. The electromagnetic potential $\mathcal{A}_\mu$ is given by,
\begin{equation}
\label{standingwave}
\mathcal{A}^{\mu}=a^{\mu}\cos\phi+b^{\mu}\sin\phi+(a'^{\mu}\cos\phi'+b'^{\mu}\sin\phi'),
\end{equation}
where $\phi=\omega t-k_z z$, $\phi'=\omega t+k_z z$, and
\begin{equation}
    a'_\mu=(0,-E_0/\omega,0,0),~~~b'_\mu=(0,0,-E_0/\omega,0).
\end{equation}
Solving the Klein-Gordon equation \eqref{K-G} in the background \eqref{standingwave}, one makes an ansatz (cf.~\eqref{a1}), 
\begin{equation}
\label{astanding}
    \psi=e^{-ip\cdot x}F(\phi,\phi').
\end{equation}
Substituting the ansatz \eqref{astanding} into the equation \eqref{K-G}, we obtain 
\begin{align}
\label{eq_KG_st}
k^2\left(\frac{\partial^2F }{\partial\phi^2}+\frac{\partial^2F }{\partial\phi'^2}\right)-2ip\cdot k\left(\frac{\partial F }{\partial\phi}+\frac{\partial F }{\partial\phi'}\right)+\\ \nonumber 
+\;(2e\;p\cdot \mathcal{A}-e^2\mathcal{A}^2)F(\phi,\phi')=0.
\end{align}
Using the subsequent ansatz, 
\begin{equation}\label{eq_F_st}
     F(\phi,\phi')=w(\zeta)e^{i\frac{(p\cdot k)}{k^2}\phi}+w(\zeta')e^{i\frac{(p\cdot k)}{k^2}\phi'},
\end{equation}
where $\zeta=\frac{1}{2}(\phi-\phi_0)$ and $\zeta'=\frac{1}{2}(\phi'-\phi_0)$, 
we finally reduce eq.~\eqref{eq_KG_st} to the system of equations,

\begin{figure}
    \centering
    \includegraphics[width=0.5\textwidth]{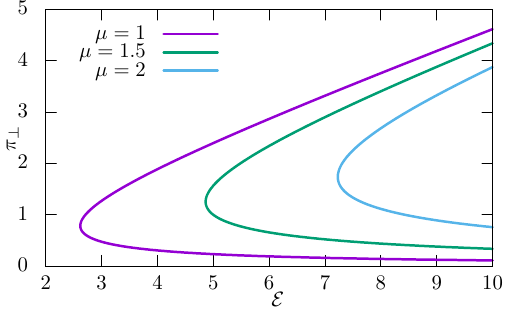}
    \caption{The boundary of broad resonance for $\pi_\perp$ as a function of $\mathcal{E}$ for fixed $1-n^2=0.1$ and different values of $\mu$. The resonance area is at the right side of the boundary.}
    \label{fig_p(e)_broad_st}
\end{figure}

\begin{figure}
    \centering
    \includegraphics[width=0.48\textwidth]{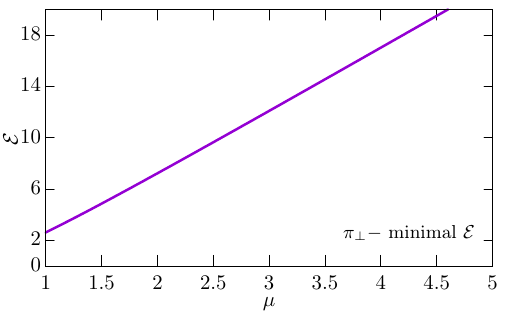}
    \caption{Dependence $\mathcal{E}(\mu)$ for the resonance boundary for fixed $1-n^2=0.1$ and $\pi_\perp$ corresponding to the minimum of $\mathcal{E}$. Broad resonance. Standing wave.}
    \label{fig_E_mu_broad_st}
\end{figure}

\begin{figure}
    \centering
    \includegraphics[width=0.99\linewidth]{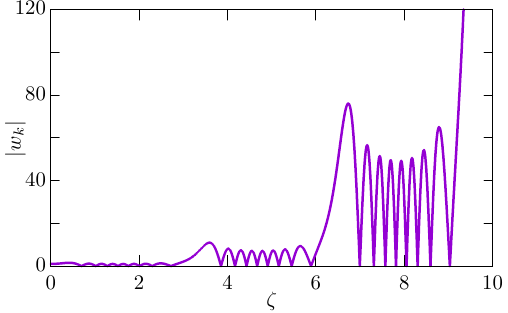}
    \caption{The growing solution of \eqref{st_Mathieu} corresponding to the broad resonance for standing wave for $\mu=1$, $\pi_\perp=0.78$ and $\mathcal{E}=3$. }
    \label{fig:sol_br_st_E3_m1_p0,78}
\end{figure}

\begin{figure}
    \centering
    \includegraphics[width=0.48\textwidth]{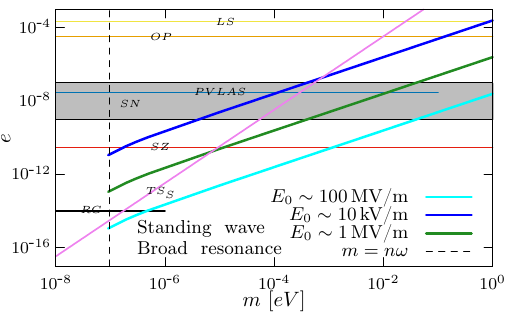}
    \caption{The projected bounds on charge $e$ and mass $m$ of mCP from the absence of anomalous losses of standing electromagnetic wave (Radio) in a metamaterial with refractive index $n=\sqrt{0.9}$, in comparison with existing bounds for mCP. Notations for experimental bounds are the same as in Fig. \ref{fig_e(m)_exp}.
    The area above the blue, green and cyan lines corresponds to the boundary of the broad resonance regime for different amplitudes  $E_0$ 
    The black dashed line corresponds to the boundary of Matheiu equation applicability $m= n\omega$.
    }
    \label{fig_e_m_exp_st}
\end{figure}

\begin{align}
\label{eq_st_before approx}
\frac{\partial^2w}{\partial\zeta^2}+\Bigg(\frac{-4e^2a^2}{k^2}(1-\cos^2(n\omega z))+ & \frac{(p\cdot k)^2}{k^4}+\\ \nonumber 
+\frac{4e}{k^2}p_{pol}\cos(n\omega z)\cos(\omega t- & \phi_0)\Bigg)w(\zeta)=0,
\end{align}
where $p_{pol}=\left((p\cdot a)^2+(p\cdot b)^2\right)^{1/2}$. The second equation of the system is eq.~\eqref{eq_st_before approx} with $\zeta$ replaced by $\zeta'$.
The equation \eqref{eq_st_before approx} is not the Mathieu equation, unlike the case of a running wave, due to its dependence on $z$. However, if the Compton wavelength of the millicharged particle is much smaller than the electromagnetic wavelength (i.e., $m \gg n\omega$), one can solve equation \eqref{eq_st_before approx} in the vicinity of an electric field antinode. By approximating the first cosine in Eq.~\eqref{eq_st_before approx} as unity, we obtain the Mathieu equation (cf.~\cite{King:2016oei}). The case of small mass requires a more detailed investigation and is not considered in this work. Introducing $\zeta_0=\frac{1}{2}(\omega t-\phi_0)$, one rewrites \eqref{eq_st_before approx} as

\begin{equation}\label{st_Mathieu}
\frac{\partial^2w}{\partial\zeta_0^2}+(A_p+2q\cos(\zeta_0))w(\zeta_0))=0,
\end{equation}
where 
\begin{equation}
A_p=\frac{(p\cdot k)^2}{k^4},
\end{equation}
\begin{equation}
q=\frac{2e}{k^2}\left((p\cdot a)^2+(p\cdot b)^2\right)^{1/2},
\end{equation}
see Appendix \ref{app_st} for details. Similar to the running wave case, using dimensionless parameters \eqref{eq:Emupi} we obtain
\begin{equation}\label{eq_Ap_st}   
A_p=\frac{\Big(\sqrt{n^2/4+\mu^2+\pi^2_\perp}-\frac{n^2}{2}\Big)^2}{(1-n^2)^2},\end{equation}
\begin{equation}\label{eq_q_st}
q=\frac{2\mathcal{E}\pi_\perp}{1-n^2}.\end{equation}
The condition for the antinode is $\mu \gg n \approx 0.95$; therefore, the narrow resonance regime cannot be considered within this approximation. However, unlike the running wave case, the broad resonance effectively occurs for the standing wave. The boundary of broad resonance $2q>A_p$ ($q\gg1$) on the configuration \eqref{eq_Ap_st}, \eqref{eq_q_st} is illustrated in Fig. \ref{fig_p(e)_broad_st} in terms of variables $\pi_\perp$, $\mathcal{E}$ for several values of $\mu$. Unlike the running wave case, the condition $q \gg 1$ is satisfied. The resonance starts for any $\pi_\perp$. Thus, for a fixed $\mu$ we determine the boundary value of $\mathcal{E}$
as an extremum of the function $\mathcal{E}(\pi_\perp)$: we solve $\left.\frac{\partial \mathcal{E}}{\partial \pi_\perp}\right|_{\pi^{0}_{\perp}}=0$ with respect to $\pi^0_{\perp}$, and take $\mathcal{E}(\pi^0_\perp)$ as the boundary value; for larger $\mathcal{E}$ the resonance starts. This boundary is illustrated in Fig.~\ref{fig_E_mu_broad_st}.

The growing solution for the broad resonance regime of the standing wave is shown in Fig. \ref{fig:sol_br_st_E3_m1_p0,78}. This solution corresponds to the point above the line in  Fig.~\ref{fig_E_mu_broad_st}, which satisfies the resonance solution. As we can see, the solution is growing in bursts, and over 10 periods, it has grown more than 100 times. The further we increase $\mathcal{E}$ in Fig. \ref{fig_E_mu_broad_st}, the faster the solution grows.


We rescale the boundary Fig.~\ref{fig_E_mu_broad_st} to the mCPs parameters for certain frequencies $\omega$ and amplitudes $E_0$ of the electromagnetic wave. For standing radio waves, we take benchmark values $\lambda\sim 1 ~$m (corresponding to a frequency $\omega\sim2\times10^{-7}$ eV),  and the different amplitudes of the electric field $E_0\sim 10\,\mbox{kV/m}$,  $1\,\mbox{MV/m}$,  and $E_0\sim 100\,\mbox{MV/m}$. The corresponding boundaries are illustrated in Fig.~\ref{fig_e_m_exp_st}, in comparison with the existing boundaries.

We now estimate the trajectory of the produced particles and the required cavity size for their confinement and accumulation. 
We consider the classical equation of motion (EOM) for counterpropagating plane waves \cite{King:2016oei, Heinzl:2008rh, Raicher:2025kpl, Lv_2021}. The EOM for a particle in the presence of an electromagnetic field is given by equation \eqref{traj_eom}.
The vector potential corresponding to the circularly polarized counterpropagating plane waves is given \eqref{standingwave}. Therefore, the trajectory of the particle for the standing wave can then be approximated as \cite{Raicher:2025kpl, Lv_2021}
\begin{equation}\label{trajectory_s}
x(t)=\frac{p_\perp t}{m}+\frac{eE_0}{\omega (k\cdot p)}\sin(\phi)+\frac{eE_0}{\omega (k'\cdot p)}\sin(\phi').
\end{equation}

Here, we are using non-relativistic expression because for standing wave we consider only massive particles.

It should be noted that, in comparison with Equation \eqref{trajectory_r}, the first term here features the mass, since the analysis concerns a massive particle and thus non-relativistic motion.

We now estimate the time the particle will travel in the perpendicular direction over a distance of $d_R\sim10\lambda\sim 10$m, using the experimental values from Fig.~\ref{fig_e_m_exp_st} (cyan line). We adopt the following values: $\omega\sim2\times10^{-7}$eV, $e\sim10^{-14}$, $E_0\sim 20 ~\mbox{eV}^2$, $\pi_\perp\sim0.8$, $m\sim10^{-6}~ $eV.  
From this calculation, we obtain the result that the first term contributes approximately $\sim 2\times 10^{-7}$ s. 

We further estimate the time required for the particle to escape the laser beam, using the laser configuration considered in section \ref{sec_exp}. For this standing wave setup, we can consider particles with a larger mass compared to the running wave case. The following laser parameters are used:  $\lambda = 1064$ nm, $\omega\sim 1.2\,\mbox{eV}$, $E\sim10^8$ eV/m and beam width $x=1\,\mbox{mm}$. The maximal electric field amplitude for the standing wave is $E_{st}=2E$. We assume the particle mass $m=1.2~$eV, which corresponds to $\mu=m/\omega=1$. From figure \ref{fig_p(e)_broad_st}, we obtain the approximate values $\pi_\perp=0.78$ and $\mathcal{E}=1.5$. Therefore, the charge is calculated as $e=\frac{\mathcal{E}\omega^2}{E_{st}}\sim5.5\times10^{-2}$.
Consequently, the characteristic timescale of the process is estimated as $t=\frac{\mu x}{\pi_\perp}\sim4.3\times10^{-12}~$s.
Note that for this laser configuration, the derived constraints are too weak for large particle masses; therefore, they are not shown in Fig.~\ref{fig_e_m_exp_st}.

\section{Conclusion}\label{sec_conclusion}

In this article, we examine the solution of the Klein-Gordon equation for scalar electrons with a tiny electric charge in a medium of refractive index $n$. We reduce the equations of motion to the Mathieu equation and investigate the instability of its solution. 
This effect is classical and occurs only for a refractive index $n<1$. Such values of $n$ can be realized in plasmas or in metamaterials.
We investigate narrow and broad resonance regimes and obtain the instability bound for mCPs.  
For the narrow resonance regime, we considered two different cases. The first case corresponds to small dimensionless energy $\mathcal{E}\ll1$, and the second corresponds to small dimensionless momentum $\pi_\perp\ll1$. Therefore, we obtain two different sets of constraints; in Figure \ref{fig_e(m)_exp}, they are shown with thick solid lines.
Consequently, we can conclude that this effect is only possible for small masses of the produced particle, which requires the condition of dimensionless mass $\mu\ll1$. 
When this condition is not satisfied, the instability boundary increases abruptly and exceeds that of the narrow resonance. Therefore, for large masses, we consider the broad resonance regime. We obtain that in this regime the instability solution is absent.

Subsequently, we compare results for narrow resonance with existing experimental constraints. In total, we obtained constraints that do not overlap with the experimental results in a limited range of parameters. In order to increase this range, one must increase the amplitude of the electric field. 

Following this, we consider the case of a standing electromagnetic background wave. We also use known solutions and boundaries of the Mathieu equation. However, in order to obtain the Mathieu equation, we use the large mass approximation. Therefore, we cannot consider small masses and the narrow resonance regime in this approximation. 
Thereafter, we investigated different values of the amplitude of the electric field for a broad resonance regime.
As a result, we obtain better constraints than in the case of a running wave. Moreover, for the radio frequency range, we estimate the required time of interaction and conclude that it is experimentally feasible. Then, we consider the Nd:YAG-laser in the infrared range and estimate the time of interaction, taking into account the laser beam width. This time corresponds to the pulse duration of this laser. 

In the next stage of our research, we plan to investigate fast radio bursts (FRBs) that interact with plasma along their propagation path \cite{Kumar:2022bob, 10.1093/mnrasl/slu039}. Moreover, if observations reveal unexplained energy losses, this could indicate the production of millicharged particles. We plan to investigate this in future work.

\paragraph{Funding} The work was supported by the grant of Russian Science Foundation № 25-22-00932.

\paragraph{Acknowledgements} The authors thank Anton Dmitriev for helpful discussions. Ekaterina Dmitrieva is a scholarship holder of the ``BASIS'' Foundation.

\appendix

\section{Solution for momentum $k-p$ of resulting particle}\label{app_k-p}

In this Appendix, we are looking for the solution of the Klein-Gordon equation \eqref{K-G} for mode $k-p$,
\[\psi_{k-p}=e^{-i(k-p)\cdot x}F_{k-p}(\phi),\]
where
\[p^2=m^2, ~\phi=k\cdot x, ~k=(\omega,0,0,n\omega).\]

Consider a circularly polarized wave \eqref{eq_polariz} and
\[\alpha=\frac{2e(p\cdot a)}{k^2},~\beta=\frac{2e(p\cdot b)}{k^2}, ~\lambda=\sqrt{\alpha^2+\beta^2}.\]
Therefore, vectors $\mathcal{A}\perp k$ and scalar product $(\mathcal{A}\cdot k)=0.$
\[2ie\mathcal{A}^\mu\partial_\mu\psi_{k-p}=(2e\mathcal{A}^\mu (k-p)_\mu F_{k-p}(\phi)+\]\[+2ie\mathcal{A}^\mu k_\mu F_{k-p}'(\phi))e^{-i(k-p)\cdot x}=2e(\mathcal{A}\cdot p) F_{k-p}(\phi)e^{-ip\cdot x}\]

As a result
\[
    k^2F_{k-p}''(\phi)-2i\,((p\cdot k)-k^2)F_{k-p}'(\phi)+\]\[+(-2e\,p\cdot \mathcal{A}-e^2\mathcal{A}^2-k^2+2(k\cdot p))F_{k-p}(\phi)=0.
\]

Setting

\[F_{k-p}(\phi)=w_{k-p}(\zeta)e^{\frac{i((k-p)\cdot k)}{k^2}\phi},\]
where $\zeta=\frac{1}{2}(\phi-\phi_0)$ and $\cos(\phi_0)=\alpha/\lambda$, $\sin(\phi_0)=\beta/\lambda$. Then $\frac{\partial \zeta}{\partial\phi}=1/2$. 



\begin{equation}
    \frac{k^2}{4}w_{k-p}''(\zeta)+(\frac{(p\cdot k)^2}{k^2}-2e(\mathcal{A}\cdot p)-e^2\mathcal{A}^2)w_{k-p}(\zeta)=0
\end{equation}

Using polarization \eqref{eq_polariz} obtain $\mathcal{A}^2=a^2$ and 
\[(\mathcal{A}\cdot p)=(p\cdot a)\cos(\phi)+(p\cdot b)\sin(\phi)=\]
\[=((p\cdot a)^2+(p\cdot b)^2)^{1/2}\cos(2\zeta),\]
we obtain the resulting equation
\[w_{k-p}''(\zeta)+\bigg[\frac{4(p\cdot k)^2}{k^4}-\frac{4e^2a^2}{k^2}-\]\[-\frac{8e}{k^2}\big((p\cdot a)^2+(p\cdot b)^2\big)^{1/2}\cos(2\zeta)\bigg]w_{k-p}(\zeta)=0.\]

We can see that for $k-p$ the equation correspond the Mathieu equation \eqref{Mathieu} for $p$ with replacement $p\to-p$,
where 
\begin{equation}
A_p=\frac{4(p\cdot k)^2}{k^4}-\frac{4e^2a^2}{k^2},
\end{equation}

\begin{equation}
q=-\frac{4e}{k^2}((p\cdot a)^2+(p\cdot b)^2)^{1/2}.
\end{equation}


\section{Floquet exponent}\label{app_Fl}

We proceed to discuss the solutions \cite{mclachlan1947theory} for the Floquet exponent, for the narrow resonance regime under the conditions $A_p\gg q,~q>0,~q-real$, $a_{ce_m}<A_p< a_{se_m}$, where $a_{cem}$ and $a_{se_m}$ are characteristic numbers of Mathieu functions and the m-number of the resonance band, $se_m$ and $ce_m$ - characteristic Mathieu functions  (sine-elliptic and cosine-elliptic) 
\begin{equation*}
w_1\simeq e^{\mu_{Fl}\zeta}[C_mce_m(\zeta,q)+S_mse_m(\zeta,q)],
\end{equation*}
\begin{equation*}
w_2\simeq e^{-\mu_{Fl}\zeta}[C_mce_m(\zeta,q)-S_mse_m(\zeta,q)],
\end{equation*}
\begin{equation}\label{eq_Fl}
    \mu_{Fl}\simeq\pm\frac{[(a_{se_m}-A_p)(A_p-a_{ce_m})]^{1/2}}{2m}.
\end{equation}
Then we consider the narrow resonance regime for the first resonance band under the conditions: $q\ll A_p,~q>0,~q-real$, $a_{ce_1}<A_p< a_{se_1}$, where $a_{ce_1}$ and $a_{se_1}$ are characteristic numbers of the Mathieu functions \cite{mclachlan1947theory, jazar2021perturbation} 
\begin{equation*}
a_{se_1}\simeq1-q-1/8q^2+\dots,
\end{equation*}
\begin{equation*}
a_{ce_1}\simeq1+q-1/8q^2+\dots
\end{equation*}
and $se_m$ and $ce_m$ characteristic functions of the Mathieu equation
\begin{equation*}
se_1=\sin(\zeta)-\frac{q}{8}\sin(3\zeta)+\dots,
\end{equation*}

\begin{equation*}
ce_1=\cos(\zeta)-\frac{q}{8}\cos(3\zeta)+\dots .
\end{equation*}
For these conditions, the expression for the Floquet exponent is
\begin{equation}
    \mu_{Fl}\simeq\frac{1}{2}\sqrt{q^2-(A_p-1)^2}.
\end{equation}
In order to begin the parametric resonance, we have the end $Re (\mu_{Fl})\neq0$, which corresponds to an exponentially growing solution. 

\section{The WKB approximation and broad resonance}\label{app_WKB}

The broad resonance corresponds to the non-perturbative limit \cite{lozanov2020reheating}. The particle production in this regime corresponds to an exponential increase solution of Hill`s equation with the Floquet exponent. The Mathieu equation is a special case of Hill`s equation with a variable frequency. For broad resonance, the particles are produced by bursts with an interval of time comparable to the oscillation period of the background. These bursts occur each time the adiabatic condition is violated \eqref{adiab_cond}, see figure \ref{fig:sol_br_st_E3_m1_p0,78}. The adiabatic condition is violated when the interaction terms vanish. The reason for the resonance is the exponential increase in occupancy number. 

We will look for a solution to equation \eqref{Mathieu} in the form of WKB solution
\begin{equation}
w_k(t)\sim\Bigg[\frac{\alpha_k}{\sqrt{2\Omega(k,\zeta)}}e^{-i\int \Omega(k,\zeta)d\zeta}+\frac{\beta_k}{\sqrt{2\Omega(k,\zeta)}}e^{i\int \Omega(k,\zeta)d\zeta}\Bigg],\end{equation}
where $|\alpha_k|^2+|\beta_k|^2=1$. The mean occupancy number $n_k=|\beta_k|^2$ in a vacuum state is equal to zero. 

In the general case  $|\alpha^j_k|^2+|\beta^j_k|^2=|\alpha^{j+1}_k|^2+|\beta^{j+1}_k|^21$, where the index $j$ corresponds to the coefficient for the $j$th and $(j+1)$th adiabatic violation.

The relation between these Bogolyubov coefficients is

\begin{equation}
\begin{pmatrix}
\alpha_k^{j+1} e^{-i\theta_k^j} \\
\beta_k^{j+1} e^{i\theta_k^j}
\end{pmatrix}
=
\begin{pmatrix}
\dfrac{1}{D_k^{j*}} & \dfrac{R_k^{j*}}{D_k^{j*}} \\[6pt]
\dfrac{R_k^j}{D_k^j} & \dfrac{1}{D_k^j}
\end{pmatrix}
\begin{pmatrix}
\alpha_k^j e^{-i\theta_k^j} \\
\beta_k^j e^{i\theta_k^j}
\end{pmatrix},
\end{equation}

where $\theta^j_k=\int^{t_j}_{t_0}\Omega(k,t)dt$ is the accumulated phase before the $j$th adiabatic violation. Coefficients $D$ and $R$ must satisfy condition $|R^j_k|^2+|D^j_k|^2=1$, in order to $\alpha$ and $\beta$ to continue being Bogolyubov coefficients. 

We assume that the initial state is vacuum $\beta^{j=0}_k=0$; therefore, after the first adiabatic violation 
\[
n_{\mathbf{k}}^{\psi,j=1} = \left|\frac{R_k^{j=1}}{D_k^{j=1}}\right|^2 .
\]

In general case
\begin{align}
n_{\mathbf{k}}^{\psi,j+1} &=
\left|\frac{R_k^j}{D_k^j}\right|^2 (n_{\mathbf{k}}^{\psi,j} + 1)
+ \left|\frac{1}{D_k^j}\right|^2 n_{\mathbf{k}}^{\psi,j} \nonumber \\
&\quad + 2\left|\frac{R_k^j}{D_k^j D_k^{j*}}\right|
\sqrt{n_{\mathbf{k}}^{\psi,j}(n_{\mathbf{k}}^{\psi,j}+1)}
\cos(\theta_k^j + \Delta\theta_k^j),
\end{align}
where
\[
\Delta\theta_k^j = \arg(R_k^j \alpha_k^j \beta_k^{j*}).
\]

In the limit $n^{\psi,j}\gg1$ we can write 
\[
n_{\mathbf{k}}^{\psi,j+1} = e^{2\mu_k^j} n_{\mathbf{k}}^{\chi,j},
\]
where $\mu_k^j$:
\begin{equation}
\mu_k^j = \ln \left|
\frac{1 + |R_k^j| e^{i(\theta_k^j + \Delta\theta_k^j)}}
{\sqrt{1 - |R_k^j|^2}}
\right|.
\end{equation}

Here, index $\psi$ corresponds to the resulting particle.

$\mu^j_{k}>0$ corresponds to non adiabatic particle production during $j$th adiabatic violation. 

For positive values of $\mu^j_k$ to occur on average, the specific functional form of $\Omega(k,t)$ must ensure that the coefficient $R^j_k $and the accumulated phase $\theta^j_k+\Delta\theta^j_k$ are properly correlated. Hence, it is possible to have stability islands even in parameter areas where the adiabaticity criterion is not met – the narrow stability strips for $A_p \lesssim2q$ in Fig. \ref{fig_Mathieu_stab/inst_Arza} (Mathieu stability chart) serve as a typical example.

\section{Expressions for standing wave}\label{app_st}

Consider equation \eqref{eq_KG_st} and substitute ansatz \eqref{eq_F_st}. The resulting equation will be
\begin{align}
\frac{\partial^2w}{\partial\zeta^2}+\Bigg(\frac{-e^2\mathcal{A}^2}{k^2}+ & \frac{(p\cdot k)^2}{k^4}+\frac{2ep\cdot\mathcal{A}}{k^2}\Bigg)w(\zeta)=0.
\end{align}

Then substitute $\mathcal{A}^\mu=a^{\mu}\cos\phi+b^{\mu}\sin\phi+a'^{\mu}\cos\phi'+b'^{\mu}\sin\phi'$ and consider each component of the equation separately. Note, that $a^2=b^2=a'^2=b'^2$, $(a\cdot b)=(a'\cdot b')=0$ and $a=-a',~ b=-b'$ and $\phi=\omega t-n\omega z$ and $\phi'=\omega t+n\omega z$.
\[\frac{e^2\mathcal{A}^2}{k^2}=\frac{e^2}{k^2}(a^2\cos^2\phi+b^2\sin^2\phi+a'^2\cos^2\phi'+b'^2\sin^2\phi'+\]\[+2aa'\cos\phi\cos\phi'+2bb'\sin\phi\sin\phi')=2a^2\frac{e^2}{k^2}(1-\cos(\phi-\phi'))=\]
\[=\frac{2a^2e^2}{k^2}(1-\cos(-2n\omega z))=\frac{4a^2e^2}{k^2}(1-\cos^2(n\omega z)).\]

Now we consider the next component with $\mathcal{A}$ and take into account that $\cos\phi_0=\alpha/\lambda$, $\sin\phi_0=\beta/\lambda$, where $\alpha=\frac{2e(p\cdot a)}{k^2}$, $\beta=\frac{2e(p\cdot b)}{k^2}$ and $\lambda=\sqrt{\alpha^2+\beta^2}$ and $\zeta=1/2(\phi-\phi_0)$, $\zeta'=1/2(\phi'-\phi_0)$.
\[\frac{2ep\cdot\mathcal{A}}{k^2}=\sqrt{(\frac{2e(p\cdot a)}{k^2})^2+(\frac{2e(p\cdot b)}{k^2})^2}(\cos(2\zeta)+\cos(2\zeta'))=\]
\[=\frac{2e}{k^2}((p\cdot a)^2+(p\cdot b)^2)^{1/2}2\cos(\omega t-\phi_0)\cos(-n\omega z)=\]
\[=\frac{4e}{k^2}((p\cdot a)^2+(p\cdot b)^2)^{1/2}\cos(\omega t-\phi_0)\cos(n\omega z).\]

As a result we obtain equation \eqref{eq_st_before approx}.

Moreover, for the case magnetic node, when $\cos(n\omega z)=1$ the expression $(1-\cos^2(n\omega z))=0$ and $\omega t-\phi_0=\zeta_0$ and we obtain the Mathieu equation \eqref{st_Mathieu}.

\bibliography{biblio}

@article{BECKER1977601,
title = {Relativistic charged particles in the field of an electromagnetic plane wave in a medium},
journal = {Physica A: Statistical Mechanics and its Applications},
volume = {87},
number = {3},
pages = {601-613},
year = {1977},
issn = {0378-4371},
doi = {https://doi.org/10.1016/0378-4371(77)90052-8},
url = {https://www.sciencedirect.com/science/article/pii/0378437177900528},
author = {W. Becker}
}

@article{CRONSTROM1977137,
title = {Photon induced relativistic band structure in dielectrics},
journal = {Physics Letters A},
volume = {60},
number = {2},
pages = {137-139},
year = {1977},
issn = {0375-9601},
doi = {https://doi.org/10.1016/0375-9601(77)90407-8},
url = {https://www.sciencedirect.com/science/article/pii/0375960177904078},
author = {C. Cronström and M. Noga}
}

@book{mclachlan1947theory,
  title={Theory and application of Mathieu functions},
  author={McLachlan, Norman William},
  year={1947},
  publisher={Clarendon Press, Oxford}
}

@book{jazar2021perturbation,
  title={Perturbation methods in science and engineering},
  author={Jazar, Reza N},
  year={2021},
  publisher={Springer}
}

@article{Kofman:1997yn,
    author = "Kofman, Lev and Linde, Andrei D. and Starobinsky, Alexei A.",
    title = "{Towards the theory of reheating after inflation}",
    eprint = "hep-ph/9704452",
    archivePrefix = "arXiv",
    reportNumber = "IFA-97-28, SU-ITP-97-18",
    doi = "10.1103/PhysRevD.56.3258",
    journal = "Phys. Rev. D",
    volume = "56",
    pages = "3258--3295",
    year = "1997"
}

@article{Gninenko:2025aek,
    author = "Gninenko, Sergei N. and Krasnikov, N. V. and Kuleshov, Sergey and Lyubovitskij, Valery E. and Crivelli, P. and Kirpichnikov, D. V. and Bueno, L. Molina and Zhevlakov, Alexey S. and Sieber, H. and Voronchikhin, I. V.",
    title = "{Probing millicharged particles with NA64$\mu$ and LDMX}",
    eprint = "2505.04295",
    archivePrefix = "arXiv",
    primaryClass = "hep-ph",
    month = "5",
    year = "2025",
    journal = ""
}

@book{lozanov2020reheating,
  title={Reheating After Inflation},
  author={Lozanov, Kaloian},
  year={2020},
  publisher={Springer}
}

@article{schmitz2010reheating,
  title={Reheating and preheating after inflation: an introduction},
  author={Schmitz, Kai and Vertongen, Gilles},
  journal={reh},
  volume={6},
  pages={t2},
  year={2010}
}

@article{our,
  title = {Tachyonic and parametric resonances for massive particle production in an intense scalar plane wave background},
  author = {Dmitrieva, Ekaterina and Satunin, Petr},
  journal = {Phys. Rev. D},
  volume = {111},
  issue = {11},
  pages = {115032},
  numpages = {11},
  year = {2025},
  month = {Jun},
  publisher = {American Physical Society},
  doi = {10.1103/8g2f-kgml},
  url = {https://link.aps.org/doi/10.1103/8g2f-kgml}
}

@article{Arza:2020zop,
    author = "Arza, Ariel",
    title = "{Production of massive bosons from the decay of a massless particle beam}",
    eprint = "2009.03870",
    archivePrefix = "arXiv",
    primaryClass = "hep-th",
    doi = "10.1103/PhysRevD.105.036004",
    journal = "Phys. Rev. D",
    volume = "105",
    number = "3",
    pages = "036004",
    year = "2022"
}

@article{PhysRev.82.664,
  title = {On Gauge Invariance and Vacuum Polarization},
  author = {Schwinger, Julian},
  journal = {Phys. Rev.},
  volume = {82},
  issue = {5},
  pages = {664--679},
  numpages = {0},
  year = {1951},
  month = {Jun},
  publisher = {American Physical Society},
  doi = {10.1103/PhysRev.82.664},
  url = {https://link.aps.org/doi/10.1103/PhysRev.82.664}
}

@article{PhysRevB.87.024408,
  title = {Broadband diamagnetism in anisotropic metamaterials},
  author = {Lapine, Mikhail and Krylova, Anastasia K. and Belov, Pavel A. and Poulton, Chris G. and McPhedran, Ross C. and Kivshar, Yuri S.},
  journal = {Phys. Rev. B},
  volume = {87},
  issue = {2},
  pages = {024408},
  numpages = {7},
  year = {2013},
  month = {Jan},
  publisher = {American Physical Society},
  doi = {10.1103/PhysRevB.87.024408},
  url = {https://link.aps.org/doi/10.1103/PhysRevB.87.024408}
}

@article{Arza:2025cou,
    author = "Arza, Ariel and Gong, Yuanlin and Shu, Jing and Wu, Lei and Yuan, Qiang and Zhu, Bin",
    title = "{Geomagnetic Signal of Millicharged Dark Matter}",
    eprint = "2501.14949",
    archivePrefix = "arXiv",
    primaryClass = "hep-ph",
    month = "1",
    year = "2025",
    journal= ""
}

@article{Berlin:2024dwg,
    author = "Berlin, Asher and Harnik, Roni and Li, Ying-Ying and Xu, Bin",
    title = "{Millicharged Condensates on Earth}",
    eprint = "2404.16094",
    archivePrefix = "arXiv",
    primaryClass = "hep-ph",
    reportNumber = "FERMILAB-PUB-24-0002-SQMS-T, USTC-ICTS/PCFT-24-08",
    month = "4",
    year = "2024",
    journal= ""
}

@mastersthesis{Lepidi:2007vnd,
    author = "Lepidi, Angela",
    title = "{Phenomenology and cosmology of millicharged particles and experimental prospects for their search}",
    eprint = "0809.4854",
    archivePrefix = "arXiv",
    primaryClass = "hep-ph",
    type = "Laurea thesis",
    school = "L'Aquila U.",
    year = "2007"
}

@article{Bogorad:2021uew,
    author = "Bogorad, Zachary and Toro, Natalia",
    title = "{Ultralight millicharged dark matter via misalignment}",
    eprint = "2112.11476",
    archivePrefix = "arXiv",
    primaryClass = "hep-ph",
    doi = "10.1007/JHEP07(2022)035",
    journal = "JHEP",
    volume = "07",
    pages = "035",
    year = "2022"
}

@article{IGNATIEV1979315,
title = {Is the electric charge conserved?},
journal = {Physics Letters B},
volume = {84},
number = {3},
pages = {315-318},
year = {1979},
issn = {0370-2693},
doi = {https://doi.org/10.1016/0370-2693(79)90048-0},
url = {https://www.sciencedirect.com/science/article/pii/0370269379900480},
author = {A.Yu. Ignatiev and V.A. Kuzmin and M.E. Shaposhnikov},
}

@article{Jaeckel:2021xyo,
    author = "Jaeckel, Joerg and Schenk, Sebastian",
    title = "{Challenging the Stability of Light Millicharged Dark Matter}",
    eprint = "2102.08394",
    archivePrefix = "arXiv",
    primaryClass = "hep-ph",
    reportNumber = "IPPP/20/81",
    doi = "10.1103/PhysRevD.103.103523",
    journal = "Phys. Rev. D",
    volume = "103",
    number = "10",
    pages = "103523",
    year = "2021"
}

@article{Holdom:1985ag,
    author = "Holdom, Bob",
    title = "{Two U(1)'s and Epsilon Charge Shifts}",
    reportNumber = "UTPT-85-30",
    doi = "10.1016/0370-2693(86)91377-8",
    journal = "Phys. Lett. B",
    volume = "166",
    pages = "196--198",
    year = "1986"
}

@article{Batell:2005wa,
    author = "Batell, Brian and Gherghetta, Tony",
    title = "{Localized U(1) gauge fields, millicharged particles, and holography}",
    eprint = "hep-ph/0512356",
    archivePrefix = "arXiv",
    reportNumber = "UMN-TH-2425-05",
    doi = "10.1103/PhysRevD.73.045016",
    journal = "Phys. Rev. D",
    volume = "73",
    pages = "045016",
    year = "2006"
}

@article{Abel:2003ue,
    author = "Abel, S. A. and Schofield, B. W.",
    title = "{Brane anti-brane kinetic mixing, millicharged particles and SUSY breaking}",
    eprint = "hep-th/0311051",
    archivePrefix = "arXiv",
    reportNumber = "IPPP-03-69, DCPT-03-138",
    doi = "10.1016/j.nuclphysb.2004.02.037",
    journal = "Nucl. Phys. B",
    volume = "685",
    pages = "150--170",
    year = "2004"
}

@article{Abel:2008ai,
    author = "Abel, S. A. and Goodsell, M. D. and Jaeckel, J. and Khoze, V. V. and Ringwald, A.",
    title = "{Kinetic Mixing of the Photon with Hidden U(1)s in String Phenomenology}",
    eprint = "0803.1449",
    archivePrefix = "arXiv",
    primaryClass = "hep-ph",
    reportNumber = "IPPP-08-14, DESY-08-026",
    doi = "10.1088/1126-6708/2008/07/124",
    journal = "JHEP",
    volume = "07",
    pages = "124",
    year = "2008"
}

@article{Shiu:2013wxa,
    author = "Shiu, Gary and Soler, Pablo and Ye, Fang",
    title = "{Milli-Charged Dark Matter in Quantum Gravity and String Theory}",
    eprint = "1302.5471",
    archivePrefix = "arXiv",
    primaryClass = "hep-th",
    reportNumber = "MAD-TH-13-01",
    doi = "10.1103/PhysRevLett.110.241304",
    journal = "Phys. Rev. Lett.",
    volume = "110",
    number = "24",
    pages = "241304",
    year = "2013"
}

@article{Brahm:1989jh,
    author = "Brahm, David E. and Hall, Lawrence J.",
    title = "{U(1)-prime DARK MATTER}",
    reportNumber = "LBL-27847, UCB-PTH-89/21",
    doi = "10.1103/PhysRevD.41.1067",
    journal = "Phys. Rev. D",
    volume = "41",
    pages = "1067",
    year = "1990"
}

@article{Feng:2009mn,
    author = "Feng, Jonathan L. and Kaplinghat, Manoj and Tu, Huitzu and Yu, Hai-Bo",
    title = "{Hidden Charged Dark Matter}",
    eprint = "0905.3039",
    archivePrefix = "arXiv",
    primaryClass = "hep-ph",
    reportNumber = "UCI-TR-2009-06",
    doi = "10.1088/1475-7516/2009/07/004",
    journal = "JCAP",
    volume = "07",
    pages = "004",
    year = "2009"
}

@article{Cline:2012is,
    author = "Cline, James M. and Liu, Zuowei and Xue, Wei",
    title = "{Millicharged Atomic Dark Matter}",
    eprint = "1201.4858",
    archivePrefix = "arXiv",
    primaryClass = "hep-ph",
    doi = "10.1103/PhysRevD.85.101302",
    journal = "Phys. Rev. D",
    volume = "85",
    pages = "101302",
    year = "2012"
}

@article{Okun:1982xi,
    author = "Okun, L. B.",
    title = "{LIMITS OF ELECTRODYNAMICS: PARAPHOTONS?}",
    reportNumber = "ITEP-48-1982",
    journal = "Sov. Phys. JETP",
    volume = "56",
    pages = "502",
    year = "1982"
}

@article{Gninenko:2006fi,
    author = "Gninenko, S. N. and Krasnikov, N. V. and Rubbia, A.",
    title = "{Search for millicharged particles in reactor neutrino experiments: A Probe of the PVLAS anomaly}",
    eprint = "hep-ph/0612203",
    archivePrefix = "arXiv",
    doi = "10.1103/PhysRevD.75.075014",
    journal = "Phys. Rev. D",
    volume = "75",
    pages = "075014",
    year = "2007"
}

@article{SENSEI:2023gie,
    author = "Barak, Liron and others",
    collaboration = "SENSEI",
    title = "{Search by the SENSEI Experiment for Millicharged Particles Produced in the NuMI Beam}",
    eprint = "2305.04964",
    archivePrefix = "arXiv",
    primaryClass = "hep-ex",
    reportNumber = "CALT-TH-2023-011, YITP-SB-2023-07, FERMILAB-PUB-23-222-PPD",
    doi = "10.1103/PhysRevLett.133.071801",
    journal = "Phys. Rev. Lett.",
    volume = "133",
    number = "7",
    pages = "071801",
    year = "2024"
}

@article{Davidson:2000hf,
    author = "Davidson, Sacha and Hannestad, Steen and Raffelt, Georg",
    title = "{Updated bounds on millicharged particles}",
    eprint = "hep-ph/0001179",
    archivePrefix = "arXiv",
    reportNumber = "CERN-TH-99-384",
    doi = "10.1088/1126-6708/2000/05/003",
    journal = "JHEP",
    volume = "05",
    pages = "003",
    year = "2000"
}

@article{Berlin:2021kcm,
    author = "Berlin, Asher and Schutz, Katelin",
    title = "{Helioscope for gravitationally bound millicharged particles}",
    eprint = "2111.01796",
    archivePrefix = "arXiv",
    primaryClass = "hep-ph",
    reportNumber = "MIT-CTP/5358, FERMILAB-PUB-21-626-T",
    doi = "10.1103/PhysRevD.105.095012",
    journal = "Phys. Rev. D",
    volume = "105",
    number = "9",
    pages = "095012",
    year = "2022"
}

@article{Chang:2018rso,
    author = "Chang, Jae Hyeok and Essig, Rouven and McDermott, Samuel D.",
    title = "{Supernova 1987A Constraints on Sub-GeV Dark Sectors, Millicharged Particles, the QCD Axion, and an Axion-like Particle}",
    eprint = "1803.00993",
    archivePrefix = "arXiv",
    primaryClass = "hep-ph",
    reportNumber = "YITP-SB-18-01, FERMILAB-PUB-17-432-T",
    doi = "10.1007/JHEP09(2018)051",
    journal = "JHEP",
    volume = "09",
    pages = "051",
    year = "2018"
}

@article{Cruz:2022otv,
    author = "Cruz, Akaxia and McQuinn, Matthew",
    title = "{Astrophysical plasma instabilities induced by long-range interacting dark matter}",
    eprint = "2202.12464",
    archivePrefix = "arXiv",
    primaryClass = "astro-ph.CO",
    doi = "10.1088/1475-7516/2023/04/028",
    journal = "JCAP",
    volume = "04",
    pages = "028",
    year = "2023"
}

@article{Heinzl:2016kzb,
    author = "Heinzl, Thomas and Ilderton, Anton and King, Ben",
    title = "{Classical and quantum particle dynamics in univariate background fields}",
    eprint = "1607.07449",
    archivePrefix = "arXiv",
    primaryClass = "hep-th",
    doi = "10.1103/PhysRevD.94.065039",
    journal = "Phys. Rev. D",
    volume = "94",
    number = "6",
    pages = "065039",
    year = "2016"
}

@article{King:2016oei,
    author = "King, B. and Hu, H.",
    title = "{Classical and quantum dynamics of a charged scalar particle in a background of two counterpropagating plane waves}",
    eprint = "1609.08105",
    archivePrefix = "arXiv",
    primaryClass = "quant-ph",
    doi = "10.1103/PhysRevD.94.125010",
    journal = "Phys. Rev. D",
    volume = "94",
    number = "12",
    pages = "125010",
    year = "2016"
}

@article{Gluck:2007ia,
    author = "Gluck, M. and Rakshit, S. and Reya, E.",
    title = "{The Lamb shift contribution of very light milli-charged fermions}",
    eprint = "hep-ph/0703140",
    archivePrefix = "arXiv",
    reportNumber = "DO-TH-07-04",
    doi = "10.1103/PhysRevD.76.091701",
    journal = "Phys. Rev. D",
    volume = "76",
    pages = "091701",
    year = "2007"
}

@article{Badertscher:2006fm,
    author = "Badertscher, A. and Crivelli, P. and Fetscher, W. and Gendotti, U. and Gninenko, S. and Postoev, V. and Rubbia, A. and Samoylenko, V. and Sillou, D.",
    title = "{An Improved Limit on Invisible Decays of Positronium}",
    eprint = "hep-ex/0609059",
    archivePrefix = "arXiv",
    doi = "10.1103/PhysRevD.75.032004",
    journal = "Phys. Rev. D",
    volume = "75",
    pages = "032004",
    year = "2007"
}

@article{DellaValle:2015xxa,
    author = "Della Valle, Federico and Ejlli, Aldo and Gastaldi, Ugo and Messineo, Giuseppe and Milotti, Edoardo and Pengo, Ruggero and Ruoso, Giuseppe and Zavattini, Guido",
    title = "{The PVLAS experiment: measuring vacuum magnetic birefringence and dichroism with a birefringent Fabry{\textendash}Perot cavity}",
    eprint = "1510.08052",
    archivePrefix = "arXiv",
    primaryClass = "physics.optics",
    doi = "10.1140/epjc/s10052-015-3869-8",
    journal = "Eur. Phys. J. C",
    volume = "76",
    number = "1",
    pages = "24",
    year = "2016"
}

@article{Burrage:2009yz,
    author = "Burrage, C. and Jaeckel, J. and Redondo, J. and Ringwald, A.",
    title = "{Late time CMB anisotropies constrain mini-charged particles}",
    eprint = "0909.0649",
    archivePrefix = "arXiv",
    primaryClass = "astro-ph.CO",
    reportNumber = "DESY-09-132, DCPT-09-124, IPPP-09-62",
    doi = "10.1088/1475-7516/2009/11/002",
    journal = "JCAP",
    volume = "11",
    pages = "002",
    year = "2009"
}

@article{Wood_2007,
doi = {10.1088/0953-8984/19/7/076208},
url = {https://dx.doi.org/10.1088/0953-8984/19/7/076208},
year = {2007},
month = {feb},
publisher = {},
volume = {19},
number = {7},
pages = {076208},
author = {Wood, B and Pendry, J B},
title = {Metamaterials at zero frequency},
journal = {Journal of Physics: Condensed Matter},
}

@article{PhysRevLett.102.093903,
  title = {Three-Dimensional Metamaterials with an Ultrahigh Effective Refractive Index over a Broad Bandwidth},
  author = {Shin, Jonghwa and Shen, Jung-Tsung and Fan, Shanhui},
  journal = {Phys. Rev. Lett.},
  volume = {102},
  issue = {9},
  pages = {093903},
  numpages = {4},
  year = {2009},
  month = {Mar},
  publisher = {American Physical Society},
  doi = {10.1103/PhysRevLett.102.093903},
  url = {https://link.aps.org/doi/10.1103/PhysRevLett.102.093903}
}

@article{navau2009magnetic,
  title={Magnetic properties of a dc metamaterial consisting of parallel square superconducting thin plates},
  author={Navau, Carles and Chen, Du-Xing and Sanchez, Alvaro and Del-Valle, Nuria},
  journal={Applied Physics Letters},
  volume={94},
  number={24},
  year={2009},
  publisher={AIP Publishing}
}

@article{magnus2008dc,
  title={A dc magnetic metamaterial},
  author={Magnus, F and Wood, B and Moore, J and Morrison, Kelly and Perkins, G and Fyson, J and Wiltshire, MCK and Caplin, D and Cohen, LF and Pendry, JB},
  journal={Nature materials},
  volume={7},
  number={4},
  pages={295--297},
  year={2008},
  publisher={Nature Publishing Group UK London}
}

@article{PhysRevB.77.092401,
  title = {Strong diamagnetic response in split-ring-resonator metamaterials: Numerical study and two-loop model},
  author = {Economou, E. N. and Koschny, Th. and Soukoulis, C. M.},
  journal = {Phys. Rev. B},
  volume = {77},
  issue = {9},
  pages = {092401},
  numpages = {4},
  year = {2008},
  month = {Mar},
  publisher = {American Physical Society},
  doi = {10.1103/PhysRevB.77.092401},
  url = {https://link.aps.org/doi/10.1103/PhysRevB.77.092401}
}

@article{Alonso-Alvarez:2019ssa,
    author = "Alonso-{\'A}lvarez, Gonzalo and Gupta, Rick S. and Jaeckel, Joerg and Spannowsky, Michael",
    title = "{On the Wondrous Stability of ALP Dark Matter}",
    eprint = "1911.07885",
    archivePrefix = "arXiv",
    primaryClass = "hep-ph",
    reportNumber = "IPPP/19/84",
    doi = "10.1088/1475-7516/2020/03/052",
    journal = "JCAP",
    volume = "03",
    pages = "052",
    year = "2020"
}

@article{baumann2011physics,
  title={The physics of inflation},
  author={Baumann, Daniel},
  journal={ICTS course},
  year={2011}
}

@article{Rubtsov:2012kb,
    author = "Rubtsov, Grigory and Satunin, Petr and Sibiryakov, Sergey",
    title = "{On calculation of cross sections in Lorentz violating theories}",
    eprint = "1204.5782",
    archivePrefix = "arXiv",
    primaryClass = "hep-ph",
    doi = "10.1103/PhysRevD.86.085012",
    journal = "Phys. Rev. D",
    volume = "86",
    pages = "085012",
    year = "2012"
}

@article{Kumar:2022bob,
    author = "Kumar, Pravir and Shannon, Ryan M. and Lower, Marcus E. and Deller, Adam T. and Prochaska, J. Xavier",
    title = "{Propagation of a fast radio burst through a birefringent relativistic plasma}",
    eprint = "2204.10816",
    archivePrefix = "arXiv",
    primaryClass = "astro-ph.HE",
    doi = "10.1103/PhysRevD.108.043009",
    journal = "Phys. Rev. D",
    volume = "108",
    number = "4",
    pages = "043009",
    year = "2023"
}

@article{10.1093/mnrasl/slu039,
    author = {Tuntsov, Artem V.},
    title = {Dense plasma dispersion of fast radio bursts},
    journal = {Monthly Notices of the Royal Astronomical Society: Letters},
    volume = {441},
    number = {1},
    pages = {L26-L30},
    year = {2014},
    month = {03},
    issn = {1745-3925},
    doi = {10.1093/mnrasl/slu039},
    url = {https://doi.org/10.1093/mnrasl/slu039},
}

@article{Davidson:1993sj,
    author = "Davidson, Sacha and Peskin, Michael E.",
    title = "{Astrophysical bounds on millicharged particles in models with a paraphoton}",
    eprint = "hep-ph/9310288",
    archivePrefix = "arXiv",
    reportNumber = "SLAC-PUB-6360, CFPA-93-TH-31",
    doi = "10.1103/PhysRevD.49.2114",
    journal = "Phys. Rev. D",
    volume = "49",
    pages = "2114--2117",
    year = "1994"
}

@article{Khlebnikov:1996mc,
    author = "Khlebnikov, S. Yu. and Tkachev, I. I.",
    title = "{Classical decay of inflaton}",
    eprint = "hep-ph/9603378",
    archivePrefix = "arXiv",
    reportNumber = "PURD-TH-96-02, OSU-TA-08-96",
    doi = "10.1103/PhysRevLett.77.219",
    journal = "Phys. Rev. Lett.",
    volume = "77",
    pages = "219--222",
    year = "1996"
}

@article{Khlebnikov:1996wr,
    author = "Khlebnikov, S. Yu. and Tkachev, I. I.",
    title = "{The Universe after inflation: The Wide resonance case}",
    eprint = "hep-ph/9608458",
    archivePrefix = "arXiv",
    reportNumber = "PURD-TH-96-06, OSU-TA-20-96",
    doi = "10.1016/S0370-2693(96)01419-0",
    journal = "Phys. Lett. B",
    volume = "390",
    pages = "80--86",
    year = "1997"
}

@article{Kofman:1994rk,
    author = "Kofman, Lev and Linde, Andrei D. and Starobinsky, Alexei A.",
    title = "{Reheating after inflation}",
    eprint = "hep-th/9405187",
    archivePrefix = "arXiv",
    reportNumber = "UH-IFA-94-35, SU-ITP-94-13, YITP-U-94-15",
    doi = "10.1103/PhysRevLett.73.3195",
    journal = "Phys. Rev. Lett.",
    volume = "73",
    pages = "3195--3198",
    year = "1994"
}

@article{Heinzl:2008rh,
    author = "Heinzl, Thomas and Ilderton, Anton",
    title = "{A Lorentz and gauge invariant measure of laser intensity}",
    eprint = "0807.1841",
    archivePrefix = "arXiv",
    primaryClass = "physics.class-ph",
    doi = "10.1016/j.optcom.2009.01.051",
    journal = "Opt. Commun.",
    volume = "282",
    pages = "1879--1883",
    year = "2009"
}

@book{Dittrich:2000zu,
    author = "Dittrich, W. and Gies, H.",
    title = "",
    doi = "10.1007/3-540-45585-X",
    isbn = "978-3-540-67428-3, 978-3-540-45585-1",
    volume = "166",
    year = "2000",
    series = "Springer Tracts in Modern Physics",
    publisher = "Springer Berlin, Heidelberg"
}

@book{Kuznetsov:2013sea,
    author = "Kuznetsov, A. and Mikheev, N.",
    title = "",
    doi = "10.1007/978-3-642-36226-2",
    isbn = "978-3-642-36225-5",
    series = "Springer Tracts in Modern Physics",
    volume = "252",
    year = "2013",
    publisher = "Springer"
}

@article{Ritus,
  title={Quantum electrodynamics phenomena in the intense field},
  author={Ritus, VI and Nikishov, AI},
  journal={Trudy FIAN},
  volume={111},
  number={5},
  year={1979},
  publisher={Nauka Moscow, Russia}
}

@article{Fedotov:2022ely,
    author = "Fedotov, A. and Ilderton, A. and Karbstein, F. and King, B. and Seipt, D. and Taya, H. and Torgrimsson, G.",
    title = "{Advances in QED with intense background fields}",
    eprint = "2203.00019",
    archivePrefix = "arXiv",
    primaryClass = "hep-ph",
    reportNumber = "RIKEN-iTHEMS-Report-22",
    doi = "10.1016/j.physrep.2023.01.003",
    journal = "Phys. Rept.",
    volume = "1010",
    pages = "1--138",
    year = "2023"
}

@article{deMontigny:2023qft,
    author = "de Montigny, Marc and Ouimet, Pierre-Philippe A. and Pinfold, James and Shaa, Ameir and Staelens, Michael",
    title = "{Minicharged Particles at Accelerators: Progress and Prospects}",
    eprint = "2307.07855",
    archivePrefix = "arXiv",
    primaryClass = "hep-ph",
    month = "7",
    year = "2023",
    journal = ""
}

@article{Fiorillo:2024upk,
    author = "Fiorillo, Damiano F. G. and Vitagliano, Edoardo",
    title = "{Self-Interacting Dark Sectors in Supernovae Can Behave as a Relativistic Fluid}",
    eprint = "2404.07714",
    archivePrefix = "arXiv",
    primaryClass = "hep-ph",
    doi = "10.1103/PhysRevLett.133.251004",
    journal = "Phys. Rev. Lett.",
    volume = "133",
    number = "25",
    pages = "251004",
    year = "2024"
}

@article{Raicher:2025kpl,
    author = "Raicher, E. and Lv, Q. Z.",
    title = "{Nonresonant quantum dynamics of a relativistic electron in counterpropagating laser beams}",
    eprint = "2502.13689",
    archivePrefix = "arXiv",
    primaryClass = "hep-ph",
    month = "2",
    year = "2025",
    journal= ""
}

@article{Lv_2021,
doi = {10.1088/1367-2630/abfa60},
url = {https://doi.org/10.1088/1367-2630/abfa60},
year = {2021},
month = {jun},
publisher = {IOP Publishing},
volume = {23},
number = {6},
pages = {065005},
author = {Lv, Q Z and Raicher, E and Keitel, C H and Hatsagortsyan, K Z},
title = {Ultrarelativistic electrons in counterpropagating laser beams},
journal = {New Journal of Physics}
}

@article{Berlin:2025btf,
    author = "Berlin, Asher and Bogorad, Zachary and Graham, Peter W. and Ramani, Harikrishnan",
    title = "{Electric Accumulation of Millicharged Particles}",
    eprint = "2510.25834",
    archivePrefix = "arXiv",
    primaryClass = "hep-ph",
    reportNumber = "FERMILAB-PUB-25-0622-SQMS-T",
    month = "10",
    year = "2025",
    journal=""
}

@article{Berlin:2025hjs,
    author = "Berlin, Asher and Bogorad, Zachary and Graham, Peter W. and Ramani, Harikrishnan",
    title = "{Cavendish Tests of Millicharged Particles}",
    eprint = "2510.25825",
    archivePrefix = "arXiv",
    primaryClass = "hep-ph",
    reportNumber = "FERMILAB-PUB-25-0623-SQMS-T",
    month = "10",
    year = "2025",
    journal = ""
}

@article{Dmitrieva:2026ges,
    author = "Dmitrieva, Ekaterina and Satunin, Petr",
    title = "{Constraints on millicharged particles from thunderstorms on the Solar system planets}",
    eprint = "2603.05338",
    archivePrefix = "arXiv",
    primaryClass = "hep-ph",
    reportNumber = "INR-TH-2026-003, NR-TH-2026-003",
    month = "3",
    year = "2026",
    journal = ""
}

\end{document}